\documentclass[showpacs,preprintnumbers,amsmath,amssymb]{revtex4}
\usepackage{graphicx}
\usepackage{url}
\begin{document}
\title{The Birth-Death-Mutation process: a new paradigm for fat tailed distributions}
\author{Yosef E. Maruvka,  David A. Kessler,  and Nadav M. Shnerb}
\affiliation{ Department of Physics, Bar Ilan University, Ramat-Gan
52900, Israel}

\begin{abstract}
Fat tailed statistics and power-laws are ubiquitous in many complex
systems. Usually the appearance of of a few anomalously successful
individuals (bio-species, investors, websites) is interpreted  as
reflecting some inherent ``quality" (fitness, talent, giftedness) as
in Darwin's theory of  natural selection. Here we adopt the
opposite, ``neutral", outlook, suggesting that the  main factor
explaining success is merely luck. The statistics emerging from the
neutral birth-death-mutation (BDM)  process is shown to fit
marvelously many empirical distributions. While previous neutral
theories have focused on the power-law tail, our theory economically
and accurately explains the entire distribution. We thus suggest the
BDM distribution as a standard neutral model: effects of fitness and
selection are to be identified by substantial deviations from it.
\end{abstract}

\pacs{ 87.18.Tt, 87.23.Kg, 05.40.-a, 87.10.Mn}

 \maketitle

Survival of the fittest or of the luckiest? The answer depends on
the subject considered. Out of ten pairs of pants bought a year ago,
the survivors are perhaps those made of a better material; if
wineglasses are considered, persistence is mainly a matter of luck.
In the absence of prior knowledge, statistics must be used in order
to identify the role of fortune: wineglass life expectancy, for
example, is described by an exponential distribution. Strong
deviations from this statistics indicate to what extent ``death" is a
result of accumulated wear, rather than from uncorrelated random
events.

In  many complex systems, though, it is hard to identify relative
role of fortune. Large differences in success (of investors or
authors) or abundance (of bio-species) do not necessarily reflect
the "quality" or the ``fitness" of the rich  and the frequent. Huge
abundance fluctuations may be a result of   accumulation of
stochastic events, as exemplified by the uneven statistics of
surnames in society~\cite{MZ}.

The schism between the ``neutral" (stochastic) and the ``fitness"
(deterministic) outlooks is most strongly pronounced in the theory
of evolutionary dynamics~\cite{raup}.  Darwin condemned those who
``attribute ... (species') proportional numbers to what we call
chance. But how false a view is this!~\cite{darwin}" and held that
the main factor shaping  eco-communities is natural selection. The
opposite view, that random drift plays the major role in evolution
\textemdash\ both on the molecular (Kimura's neutral
evolution~\cite{kimura}) and the ecological (Hubbell's community
drift model~\cite{hubbell}) levels \textemdash\  has sparked a
series of ongoing hot and emotional debates.

In economy and social sciences the deterministic approaches tend to
emphasize the tremendous inequality in income and wealth, say, as
reflecting underlying ``quality" (from prudence to crookedness)
differences. The opposing neutral approach~\cite{gibrat} have
recently found a prominent outspoken, Nassim Taleb. In his
books~\cite{taleb1,taleb2} he maintains that the weight of
unpredictable events (what he calls ``black swans") is overwhelming in determining
economic and social success.

Purely deterministic and purely stochastic theories are both
oversimplifications. The real scientific problem  is to find  the
relative weight of chance  versus fitness.  The assumption of
neutral dynamics is most useful as a null hypothesis, with which
empirical statistics should be compared. Nowadays this role is
played by the Yule-Simon statistics~\cite{yule,simon,newman1}, or
its approximation by  a simple power law.  In the following we
briefly review Yule's model and point out its major shortcoming. We
suggest a correction that yields different statistics and  show that
the new distribution fits many ``canonical" empirical datasets very
nicely.

Yule-Simon  theory~\cite{yule} arose from a study of the   the
highly skewed distribution of biological species within genera. One
of the graphs studied by Yule \textemdash\  for the family of
long-horn beatles Cerambycinea \textemdash\ is plotted
 in the left inset of \verb+Fig. 1+. This is a Pareto plot showing $n_m$,
 the fraction of genera with $m$ species, vs. $m$ on a
log-log scale. One observes a few ``wealthy" genera to which many
species belong, and many ``poor" genera with apparent linear
dependence that suggests a power-law distribution.

Yule's neutral model posited that the rate of speciation is fixed
for all species. Upon speciation, the new species stays in  the same
genus with probability $1-\mu$.   $\mu$, the chance  that the
offspring species branches out to form a new genus, is also fixed,
ensuring perfect neutrality (no fitness). This simple process
generates a steady state distribution that converges rapidly to a
power law for the relative species abundance $n_m$,
\begin{equation} \label{1}
n_m = C B(m,2+\mu) \sim C m^{-(2+\mu)}.
\end{equation}
where $C$ is a normalization factor. Note that this fat-tailed
distribution has nothing to do with the ``quality" differences among
species, instead it is a result of the multiplicative character of
the noise.

As pointed out by Herbert Simon~\cite{simon}, Yule's argument goes
far beyond its original context. Simon considered  power-laws for
the number of occurrences of words in a text, scientific
publications  and wealth distribution. Subsequently, the appearance
of power-laws has been recognized as a fundamental feature of eco-,
econo-, bio- and socio-systems, with countless of examples from
protein family statistics~\cite{Shakhnovich}, surname abundance
ratio~\cite{MZ,jtb}, internet connections~\cite{barabasi},
 firm sizes~\cite{simon58},
casualties in terror attacks~\cite{terror1}   and so on. In addition
the common scenario considered in the new popular theory of scale
free networks  - the preferential attachment dynamics - is indeed
mathematically equivalent to Yule's  process [see Appendix A] where
small families are generated by a source, not by
mutations~\cite{newman1}.

As a starting point for the presentation of our new neutral model,
let us stick for the moment to the original context of Yule theory,
 the species within genera statistics. The main panel of
\verb+Figure 1+ reveals a major failure of the Yule-Simon model. The
original distribution observed by Yule for  Cerambycinea beatles,
based on the 1024 genera (5719 species) is compared with the current
data with 27221 species and 4411 genera. Clearly, something bad has
happened to the simple power-law: it characterizes now only the tail
of the distribution, and a very pronounced ``shoulder" appears for
the small genera.

\begin{figure}
\centering
\begin{center}
\hspace{0cm}
\includegraphics[width=8cm]{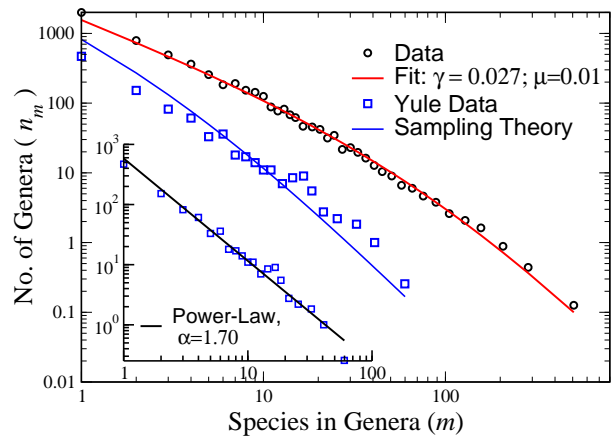}
\end{center}
\caption{Species within genera statistics for Cerambycinea beatles.
The original species within genera statistics used by Yule (blue
squares), based on 1024 genera known at 1925 for the Cerambycinea
family (down left). On a log-log scale this graph looks very much
straight, suggesting a power-law statistics (black line). In the
main figure, the black circles show the contemporary statistics as
obtained for 4411 genera (27221 species of Cerambycinea~\cite{COL}),
where a pronounced ``shoulder" appears. The red line is the best fit
of BDM function (\ref{power}), $\gamma$ is the diversification rate
and $\mu$ reflects the chance of a new species to initiate a new
genus. The blue line shows the prediction of our theory for a sample
of $R_0 = 5719$ species chosen at random out of the 22271 known
today with the same $\mu$ and $\gamma$, as obtained from Eq.
\ref{sample}. This is now a prediction \emph{without any fitting
parameters}, to be compared with the original Yule statistics.  }
\end{figure}

This shoulder appears in almost any fat-tailed
distribution~\cite{newman1}. Accordingly, a ``power law fit"
indeed involves  \emph{two parameters}: a threshold $x_{min}$
marking the end of the shoulder and the tail's slope. Unfortunately,
the large argument tail  tends to be of poor quality, noisy, brutish
and short. Very rarely one finds a reliable dataset that  allows for
a good quality fit. Indeed, a recent metaanalysis by Clauset,
Shalizi and Newman~\cite{newman2} reveals that, among 20 canonical
datasets considered, only in one case  a power law  fit is really
convincing and in most  cases other  distributions are doing better.

We suggest that these  obstacles reflect an essential shortcoming of
the Yule-Simon theory: the neglect of ``death"  events. In reality
species go extinct, individuals die and links break down, yet in the
Yule-Simon theory this never happens. A death process cannot be
taken into account by simply introducing a net birth rate; it also
accounts for the stochastic extinction of existing families
(genera). Yule theory thus overestimates the fraction of small
families, which explains the  typical ``shoulder" that appears at small
$m$'s.

ƒ
\begin{figure*}
\begin{center}
\hspace{0cm}
\includegraphics[width=13cm]{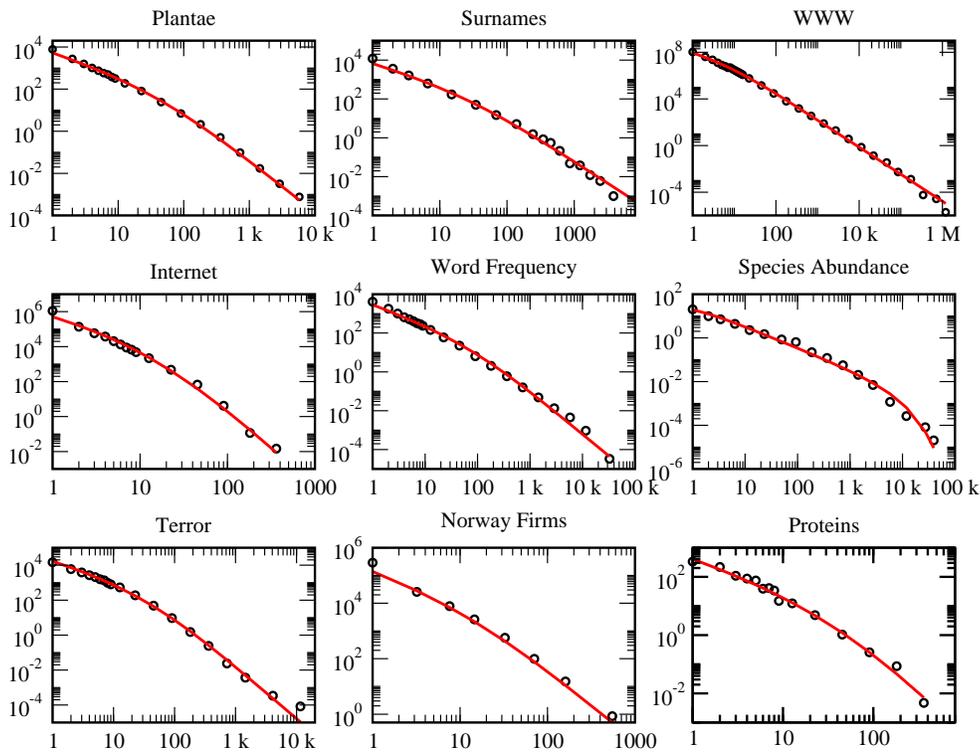}
\end{center}
\caption{Tour de force of BDM statistics: Pareto plots are presented
for empirical datasets obtained from independent studies across many
disciplines. The best fit values of $\gamma$ and $\mu$ are given for
each item.  (a) Species within genera statistics for the Plantae
kingdom~\cite{COL}  $\gamma=0.055  \  \mu=0.017$. (b) Surname
statistics from the 1790 US census. The growth rate ($\gamma =
0.034$) was inferred~\cite{jtb} from historical censuses in England,
and the fit retrieves the ``mutation" (surname changes) rate to be
$\mu=0.011$. (c) WWW: number of sites with certain degree of links
as a function of the degree. The set of 200 million web pages with
1,500 million hyperlinks first considered by Broder et.
al.~\cite{Broder} has been analyzed. $\gamma=0.27 \ \mu= 0.065$. (d)
Internet (physical structure) - number of nodes with $m$ links vs.
m. Data obtained from DIMES web site (www.netdimes.org).
$\gamma=0.72 \ \mu=0.51$. (e) Word frequency, number of words that
occur $m$ times in the King James Bible (KJB)
(http://www.htmlbible.com). $\gamma=0.021 \ \mu=0.004$. (f) Species
abundance ratio in the tropical forest~\cite{hubbella}. Here $\gamma
= 5.4 \cdot 10^{-5} \ \mu = 1.5 \cdot 10^{-4}$. (g) Human
insurgency: number of terror attacks with $m$ casualties vs. $m$.
Data from Global Terrorism Database, START
(http://www.start.umd.edu).  $\gamma=0.1 \ \mu=0.051$. (h) Number of
Norwegian firms with $m$ employees, as obtained from statistics
Norway website,  www.ssb.no. (Data for 2010). $\gamma=0.11 \
\mu=0.04$. (i) Domain family size distribution for Drosophila
melanogaster~\cite{Karev}. $\gamma=0.087 \ \mu=0.046$.   }
\end{figure*}

Recently Manrubia and Zannete~\cite{MZ} studied the distribution of
surnames in a population, using a model which is a specific example
of the birth-death-mutation (BDM) process (see also \cite{maritan}).
We~\cite{jtb} then extended these results, showing that the
resulting distribution is independent of the particular details of
the process.  In the spirit of Simon's realization that the Yule
model results are applicable in a much broader context, we here
propose, and demonstrate by numerous examples, that the BDM process
and its resulting statistics should be applicable to a very wide
range of empirical datasets.

Here is a list of the main results for  the  statistics of the BDM
process, where the total population is growing/decaying at rate
$\gamma$.  In the  supplementary material we resent a detailed
description of the BDM dynamics and  establish the equivalence
between this process and preferential attachment~\cite{barabasi}
with the possibility of link removal.

\begin{enumerate}
  \item The probability distribution function (the chance $n_m$ to pick at random a family of size $m$)
  is described by the Kummer function $U(a,b,c)$ \cite{Abramowitz}.

\begin{enumerate}
\item If the growth rate   $\gamma$ is larger than the mutation rate
$\mu$, an asymptotic power-law tail appears:
\begin{eqnarray} \label{power}
n(m)  &=& \frac{\nu R_c \Gamma(2+\nu)}{m}\, U\!\left(1+\nu,0,\frac{R_c
m}{N_0}\right)  \nonumber\\
&\ &\qquad\qquad \stackrel{{\tiny{ m\to\infty}}} {\sim} \  m^{- (
1+\frac{\gamma}{\gamma-\mu} ) },\nonumber
\end{eqnarray}
where $\nu \equiv \mu/(\gamma - \mu)$  and $R_c \equiv 2 N_0 |\gamma
- \mu|/\sigma^2$, $N_0$ is the current population size.

\item For $\mu > \gamma$, the BDM dynamics supports a truncated
  power-law distribution [here $\nu \equiv \gamma/(\mu-\gamma)$],
\begin{eqnarray} \label{trunk}
n(m) &=& \frac{ R_c \Gamma(1+\nu)}{m}\, U\!\left(\nu,0,\frac{R_c
m}{N_0}\right) e^{-\frac{R_c m}{N_0}} \nonumber\\
&\ &\qquad\qquad \stackrel{{\tiny
m\to\infty}}{\sim}\ m^{-1- \nu}
e^{-\frac{2}{\sigma^2}(\mu-\gamma)m}.
\end{eqnarray}

\end{enumerate}

\item When $R_0$ individuals are sampled the effective strength of the
sampling is $s = R_0/R_c$. 
In the strong sampling limit, $s \gg 1$,
the new distribution is just a rescaled Kummer~\cite{jtb}. On the
other hand if $s \ll 1$,
\begin{equation}\label{sample}
n^R(m) \approx \textrm{B}(m-1-\nu,2+\nu)\nu R_o s^\nu.
\end{equation}
\end{enumerate}

Eq. (\ref{sample}) implies that the BDM statistics crosses over to
the Yule-Simon result when the sampling is weak [see Eq. (\ref{1})
and the discussion in Appendix B].  Since weak sampling yields
mainly members of large families for which the chance of extinction
is small, Yule's theory with a net birth rate becomes adequate.
Indeed, in the main part of \verb+Fig. 1+ we show how the BDM Kummer
statistics fits the contemporary data for Cerambycinea and how one
can reconcile the Yule result by taking into account the effect of
sampling. Note that our theory~\cite{jtb} is based on a
Fokker-Planck equation that fails when the size of the family is of
order unity~\cite{WKB}, thus here and in the following figures the
curve fails to fit the number of singletons.

\verb+Fig. 2+  demonstrates the power of our technique using many
paradigmatic fat-tailed distributions from the social sciences
(surnames, insurgency, WWW), engineering (internet),   ecology
(species within genera, species abundance ratio), linguistics (word
frequency) biology (protein family statistics) and economy (firms
size distribution). In all cases presented here a two parameter fit
is shown, thus we are not using more fitting parameters than a
standard power-law fit. In some cases the relevance of the BDM
dynamics to the underlying process is clear; for example, it is very
close to the birth-death-innovation process already suggested for
proteins~\cite{Karev}. In other cases (terror attacks, word
frequency) the underlying process is not well understood, and more
studies are needed in order to prove, or disprove, the relevance of
BDM, perhaps along the lines suggested by \cite{terror, vito}. The
agreement of theory and data is impressing with respect to other
fits on log-log scale; some examples of other fitting functions and
 distributions are given in Appendix D.

Clearly the BDM theory is much stronger than a simple power-law fit,
yielding sharper predictions and fitting almost perfectly many
paradigmatic empirical datasets. Its amazing success, even where the
BDM process is certainly   a crude approximation for the real
dynamics, suggests that this distribution behaves like a central
limit for many multiplicative neutral processes.

\begin{figure}
\centering
\includegraphics[width=8cm]{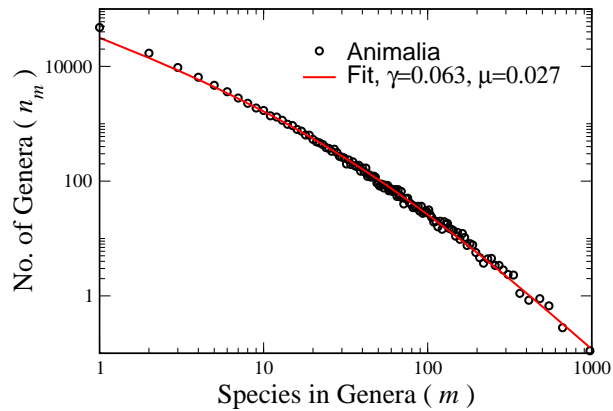}
\caption{  A Pareto plot for the species within genera statistics
for the Animalia kingdom. The fit of the BDM theory to the data is
surprisingly good, given the existence of different taxonomical
classifications for genera.  The fit suggests a diversification
(speciation minus extinction) rate of about 0.063; this value falls
within the confidence intervals obtained by Ricklefs~\cite{ricklefs}
for North and South American clades of passerine birds. }
\end{figure}

For any of the topics of \verb+Fig. 2+  a comprehensive discussion
is needed in order to put our new results for $\gamma$ and $\mu$ in
the context of the specific field.  This is beyond the scope of this
Letter, and short specific  comments are presented in Appendix C.

Let us conclude by demonstrating  the quality of our results using
one example. \verb+Figure 3+ shows the species within genera
statistics for all the Animalia kingdom~\cite{COL}. The Kummer function
fits almost exactly the empirical data, much better than other
distributions conjectured (see SM). The rate of diversification
(speciation minus extinction), $\gamma = 0.063 \pm 0.02$, is
consistent with the range of values estimated from lineage through
time plots~\cite{ricklefs}, and our confidence intervals are much
tighter.

\acknowledgements{ This work  was supported by the CO3 STREP
of the Complexity Pathfinder of NEST (EC FP6). We thank Robert
Ricklefs and  David Aldous for helpful comments and discussions.}

\appendix

\section{ The birth-death-mutation process}

The birth-death-mutation (BDM) process, in its simplest form,
governs the dynamics of $S$ families of agents. Each family is
characterized by $m$, the number of agents in it. For the sake of
concreteness let us consider a population of species (agents), each
of which belongs to a genus (family).

At every time step a species is chosen at random among all species,
independent of its genus. This agent is removed with probability
$1-p$ and reproduces (speciates) with probability $p$. The offspring
belongs to the same genus  as its parent species with probability
$1-\mu$, and "mutates" to form a new genus  with probability $\mu$.
Note that we use the word  "mutation" to indicate an offspring that
forms a new family (genus, surname,protein family), rather than
belonging to the same clan as its parent. The parameter $\gamma =
2p-1$ defines the growth rate (if positive) or the decay rate (if
negative) of the population. This is the overlapping generations
(Moran) version of the process.

Many other processes support the same steady state distribution of
family sizes~\cite{jtb}. Of particular importance is the
nonoverlapping generations (Wright-Fisher) version of this dynamics.
In this case all agents produces offspring at once and then are
removed. An agent produce $n$ offspring with probability $P_n$. The
average number of offspring per individual is thus given by $\bar{n}
= \sum n P_n$, and the growth/decay rate is $\gamma = \bar{n} - 1$.
Again $\mu$ is the mutation rate as described above.

In previous work~\cite{jtb} we have shown that all these processes
yield the same steady-state distribution of family sizes, which is
independent of the "microscopic" details. The final distribution
depends only on the growth rate $\gamma$,  the mutation rate $\mu$,
and the variance $\sigma^2 = Var(n)$. For the Moran case $\sigma^2
=2$. It turns out that $n(m)$ satisfies the Kummer differential
equation
\begin{equation}
\frac{\partial n(m)}{\partial t} = \frac{\sigma^2}{2}
\frac{\partial^2}{\partial m^2} \left[ mn(m) \right] + (\mu-\gamma)
\frac{\partial}{\partial m} \left[ mn(m) \right].
\end{equation}
Note that this equation resembles a    diffusion-convection process
for $mn(m)$.

The same statistics emerges if agents are removed with probability
$1-p$, reproduce into the parent set with probability $p(1-\mu)$,
and new agents, each deposited into an empty set (family), are added
with probability $p \mu$ (we refer to this as the birth-death-source
process, BDS). This is the case, e.g.,  if nodes, each carrying  a
certain number of links, are added to an already existing network
and the chance of a link to be attached to an already existing node
is proportional to the degree of the node. If links are removed at a
different rate, the process yields the same statistics as the BDM
(up to slight modifications since new families appear, in realistic
networks, with size which is greater than one).

The BDM process is a generalization of the famous Yule process which
has no death in it; i.e., agents are only born and  mutate. In the
same sense, the BDS version generalizes the preferential attachment
process~\cite{barabasi} in which links are only added to the network
but are never removed.

\section{Yule-Simon statistics as a weak sampling limit of BDM}

In the process defined by Yule there is no death, and the mutation
rate $\mu_{\textit{\tiny  Yule}}$ is simply the ratio between the
average number of new surnames (or genera) that appear during a
period of time and the number of new individuals added, during the
same period, to already existing families (see the detailed
discussion in~\cite{newman1}).

In the BDM process the rate in which new families are generated is
$\mu b N$ ($N$ is the total population at certain time, $b$ is the
birth rate) and the rate in which the total population in the
already existing families grows is $[b(1-\mu) - d]N$. Without loss
of generality we can choose $d=1$ such that $b = 1+\gamma$, since
the growth rate $\gamma \equiv b-d$. The ratio between the new
families generation rate and the old families growth rate is, (to
the first order in the small parameters $\gamma$ and $\mu$), $ \nu
\equiv \mu / (\gamma - \mu)$. This implies that for small growth and
mutation rates, which is the regime of validity of the Kummer
theory, Yule theory is equivalent to BDM iff stochastic extinction
is neglected and $\mu_\textit{\tiny Yule} $ is replaced by $\nu$.
For that reason, Eq. (5) of the main text is equivalent to Yule
statistics (Eq. 1) with $\nu$ instead of $\mu$.

\section{Remarks for Figure 2 of the main text}

The remarks below refer to the panels of Fig. 2:

General: The binning of the data was done using  a half logarithmic
scale, which means that for small families ($m \le 10$) we had a bin
for every number, while for large families we used logarithmic
binning with a bin size  $2^k$ ($k$ is the bin number).  We have
found this to be optimal in terms of presentation clarity, but the
Kummer fit has been checked using other binning schemes and the
differences are negligible.  For two datasets (surname panel (b),
and firms panel (h)) the data was available only in a  binned form,
so the existing binning  scheme has been retained.

\begin{description}
  \item[(a)] The statistics of the Plantae kingdom. This dataset is
  similar to the Animalia displayed and analyzed in Fig. 3; we have
  preferred to present a more detailed analysis of Animalia since this
  is the largest kingdom.
  \item[(b)] Surname: The size of a family was defined as the number of households
  having the same
  surname. Data refer to the US census of 1790, when the US
  population shared the same genealogic and demographic histories with the British population.
  The English demography is roughly documented since the Domesday Book  census carried out by William the Conquerer. For more details
  see~\cite{jtb}.
  \item[(c)] WWW links statistics. There is some ambiguity about the
  kind of sampling involved in the collection of the data. In
  principle one should make a distinction between  building a surname
  statistics by sampling \emph{individuals} and asking for their surname,
  in which case Eq. (5) of the method section is applicable, and
  sampling surnames and asking for the number of individuals having
  this specific surname. In the internet case the sampling is done
  by crawlers moving from node to node along the links; here a link is
  an individual and a node is a "surname". In any case,
the success of our fit to a full census theory means that the effect
of sampling, if any, is weak (i.e., that we are in the strong
sampling regime).

\item[(d)]  We present here the nodes in-degree distribution
(i.e. the size of a node is determined by the number of links
pointing to it). The nodes out-degree distribution does not follow
 Kummer. This difference needs further analysis.

  \item[(e)] The counting of the words was done using StatA \cite{Sta}.
   We should mention that different tools define words in slightly different
   ways;
   however the distributions produced by the different tools are still almost identical.
\item[(f)] The data was averaged over six different censuses. Time
between consecutive censuses is five years, to be compared with the
lifetime of a tree which is typically about 100 years.

Our best fit yields $\gamma = 4.3 10^{-5}$ and  $\mu  = 2.9
10^{-4}$. This suggests that the total population of the
meta-community isn't really fixed but rather grows extremely slowly.
Although the model is neutral, the overall effect of adaptation may
very slowly increase the carrying capacity of the forest.

While we are not trying to claim that our fit is actually
conclusive, this result opens an interesting possibility for
refutation of the critics of the "point mutation" version of
Hubbell's theory, who base themselves on turnover rates. As pointed
out by Ricklefs~\cite{ricklefs2006} and by Nee~\cite{Nee} the time
to origination of a species with N individuals is about 2N. This
leads to ridiculously large timescales when applied to realistic
species abundance.  One implication of our work is that the
introduction of a very weak growth rate does not kill the
statistics, yet it clearly shortens the time to origination
significantly. For example for 10 million trees with generation time
of a 100 years, the time to origination if the total population is
fixed will be of order of a billion years, while for the  $\gamma$
above it will be 40 million years.

  \item[(g)] The datasets had also some non-integers values (the meaning of which is unclear to us)
   that we rounded  up to the closest integer number.

\item[(h)] The dataset includes the  number
of establishments with $m$ employees, starting from $m=0$. In order
to avoid this zero we have shifted $m \to m+1$, counting the owner
also as an employee.

\item[(i)] The source of data is Fig. 7b of Karev et.
  al.\cite{Karev} whereas our presentation uses   logarithmic binning.
  Our birth-death-mutation process differs from the model suggested
  by Zeldovich et. al.\cite{Shakhnovich} that does not include
  "death" (proteins never disappear from the system). The birth-death-innovation model
  suggested by Karev et. al.~\cite{Karev} admits death, but the
  innovation rate (the rate in which new protein families are
  generated) is independent of the population size, and the birth
  rate depends on the size of the family. Thus this model is not
  really neutral.

\end{description}

\section{The adequacy of Kummer}

When dealing with fat-tailed distributions that are extended over
many orders of magnitude, a log-log plot must be used. However,
these plots are notoriously known to smear out some fine details of
the distribution, and sometimes this feature blurs the actual
mismatch between the theory and the empirical data. The level of
exactness is thus a crucial factor in determining  the adequacy  of
a fit. Here we describe two examples.

First, in Fig. \ref{1} the Kummer best fit is compared with the best
fit obtained for the modified Pareto (Zipf - Mandelbrot)
distribution, which is a two parameter law with the same concave
shape,
\begin{equation}
n_m = N_0 \frac{(a+1)^{b-1}(b-1)}{(a+m)^b}
\end{equation}
where $N_0$ is the population size. The best fit for the parameters
$a$ and $b$ is shown together with the best Kummer fit. One can see
that, although the mismatch is never large in a loglog plot once the
function captures the general trend, there are systematic deviations
in the modified Pareto case but not from the Kummer function (note
again that the singletons are not covered by our theory so the
mismatch at $m=1$ is irrelevant).

\begin{figure} \label{1}
\centering
\includegraphics[width=8cm]{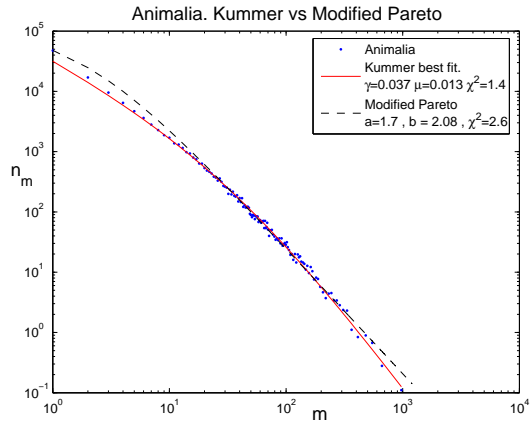}\\
\caption{\textbf{Animalia kingdoms statistics:} Modified pareto
(Zipf-Mandelbrot, dashed line) best fit vs. Kummer best fit.}
\end{figure}

\begin{figure} \label{2}
\centering
\includegraphics[width=8cm]{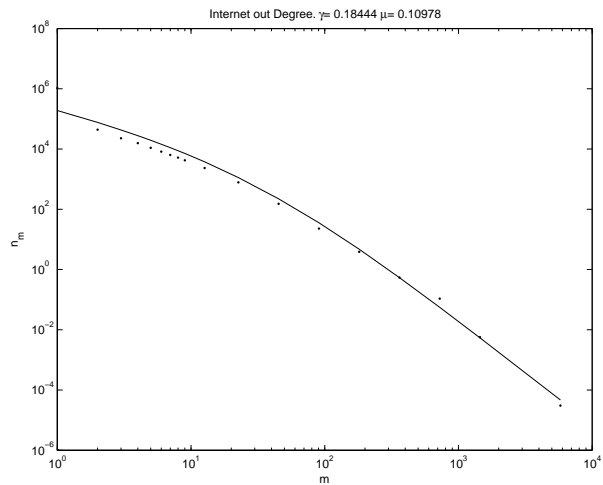}\\
\caption{\textbf{Out-degree statistics:} The best fit to Kummer
fails systematically at small $m$s. }
\end{figure}

As another example let us present a case where  systematic
deviations from Kummer show up. In Fig. \ref{2} the  out-degree
distribution of nodes in the internet (the in-degree that satisfies
Kummer is shown in Fig. 2d) is shown together with the best fit to
Kummer, and indeed one can see systematic deviations that makes the
Kummer fit  very suspicious, if not fully disqualified.

In general the Kummer function may be considered in any case where
the distribution is monotonically decreasing (so it is inappropriate
as an explanation to, say, scientific citation statistics where a
hump appears at intermediate values of $m$). For a reasonable fit
the  slope at small $m$-s should be close to one, not too shallow
(as in the Tsallis distribution~\cite{tsallis}) or too steep.


\begin{thebibliography}{27}
\expandafter\ifx\csname
natexlab\endcsname\relax\def\natexlab#1{#1}\fi
\expandafter\ifx\csname bibnamefont\endcsname\relax
  \def\bibnamefont#1{#1}\fi
\expandafter\ifx\csname bibfnamefont\endcsname\relax
  \def\bibfnamefont#1{#1}\fi
\expandafter\ifx\csname citenamefont\endcsname\relax
  \def\citenamefont#1{#1}\fi
\expandafter\ifx\csname url\endcsname\relax
  \def\url#1{\texttt{#1}}\fi
\expandafter\ifx\csname urlprefix\endcsname\relax\def\urlprefix{URL
}\fi \providecommand{\bibinfo}[2]{#2}
\providecommand{\eprint}[2][]{\url{#2}}

\bibitem[{\citenamefont{Manrubia and Zanette}(2002)}]{MZ}
\bibinfo{author}{\bibfnamefont{S.~C.} \bibnamefont{Manrubia}} \bibnamefont{and}
  \bibinfo{author}{\bibfnamefont{D.~H.} \bibnamefont{Zanette}},
  \bibinfo{journal}{J. Theor. Biol.} \textbf{\bibinfo{volume}{216}},
  \bibinfo{pages}{461} (\bibinfo{year}{2002}).

\bibitem[{\citenamefont{Raup}(1992)}]{raup}
\bibinfo{author}{\bibfnamefont{D.~M.} \bibnamefont{Raup}},
  \emph{\bibinfo{title}{Extinction: bad genes or bad luck}}
  (\bibinfo{publisher}{WW Norton \& Company}, \bibinfo{address}{New York},
  \bibinfo{year}{1992}).

\bibitem[{\citenamefont{Darwin}(1859)}]{darwin}
\bibinfo{author}{\bibfnamefont{C.}~\bibnamefont{Darwin}},
  \emph{\bibinfo{title}{The origin of species by means of natural selections}}
  (\bibinfo{publisher}{John Murrary}, \bibinfo{address}{London},
  \bibinfo{year}{1859}).

\bibitem[{\citenamefont{Kimura}(1985)}]{kimura}
\bibinfo{author}{\bibfnamefont{M.}~\bibnamefont{Kimura}},
  \emph{\bibinfo{title}{The neutral theory of molecular evolution}}
  (\bibinfo{publisher}{Cambridge Univ. Press}, \bibinfo{address}{Cambridge},
  \bibinfo{year}{1985}).

\bibitem[{\citenamefont{Hubbell}(2001)}]{hubbell}
\bibinfo{author}{\bibfnamefont{S.~P.} \bibnamefont{Hubbell}},
  \emph{\bibinfo{title}{The unified neutral theory of biodiversity and
  biogeography}} (\bibinfo{publisher}{Princeton Univ. Press},
  \bibinfo{address}{Princeton, NJ}, \bibinfo{year}{2001}).

\bibitem[{\citenamefont{Gibrat}(1931)}]{gibrat}
\bibinfo{author}{\bibfnamefont{R.}~\bibnamefont{Gibrat}},
  \emph{\bibinfo{title}{Les in{\'e}galit{\'e}s {\'e}conomiques}}
  (\bibinfo{publisher}{Librairie du Recueil Sirey}, \bibinfo{year}{1931}).

\bibitem[{\citenamefont{Taleb}(2005)}]{taleb1}
\bibinfo{author}{\bibfnamefont{N.}~\bibnamefont{Taleb}},
  \emph{\bibinfo{title}{Fooled by randomness: The hidden role of chance in the
  markets and life}} (\bibinfo{publisher}{Random House}, \bibinfo{address}{New
  York}, \bibinfo{year}{2005}).

\bibitem[{\citenamefont{Taleb}(2007)}]{taleb2}
\bibinfo{author}{\bibfnamefont{N.}~\bibnamefont{Taleb}},
  \emph{\bibinfo{title}{The Black Swan: The Impact of the Highly Improbable}}
  (\bibinfo{publisher}{Random House}, \bibinfo{address}{New York},
  \bibinfo{year}{2007}).

\bibitem[{\citenamefont{Yule}(1925)}]{yule}
\bibinfo{author}{\bibfnamefont{G.~U.} \bibnamefont{Yule}},
  \bibinfo{journal}{Phil. Trans. R. Soc. Lond. B}
  \textbf{\bibinfo{volume}{213}}, \bibinfo{pages}{21} (\bibinfo{year}{1925}).

\bibitem[{\citenamefont{Simon}(1955)}]{simon}
\bibinfo{author}{\bibfnamefont{H.}~\bibnamefont{Simon}},
  \bibinfo{journal}{Biometrika} \textbf{\bibinfo{volume}{42}},
  \bibinfo{pages}{425} (\bibinfo{year}{1955}).

\bibitem[{\citenamefont{Newman}(2005)}]{newman1}
\bibinfo{author}{\bibfnamefont{M.~E.~J.} \bibnamefont{Newman}},
  \bibinfo{journal}{Contemporary Phys.} \textbf{\bibinfo{volume}{46}},
  \bibinfo{pages}{323} (\bibinfo{year}{2005}).

\bibitem[{\citenamefont{Zeldovich et~al.}(2007)\citenamefont{Zeldovich, Chen,
  Shakhnovich, and Shakhnovich}}]{Shakhnovich}
\bibinfo{author}{\bibfnamefont{K.~B.} \bibnamefont{Zeldovich}},
  \bibinfo{author}{\bibfnamefont{P.}~\bibnamefont{Chen}},
  \bibinfo{author}{\bibfnamefont{B.~E.} \bibnamefont{Shakhnovich}},
  \bibnamefont{and} \bibinfo{author}{\bibfnamefont{E.~I.}
  \bibnamefont{Shakhnovich}}, \bibinfo{journal}{PLoS Comp. Biol.}
  \textbf{\bibinfo{volume}{3}}, \bibinfo{pages}{e139: 1224}
  (\bibinfo{year}{2007}).

\bibitem[{\citenamefont{Maruvka et~al.}(2009)\citenamefont{Maruvka, Shnerb, and
  Kessler}}]{jtb}
\bibinfo{author}{\bibfnamefont{Y.~E.} \bibnamefont{Maruvka}},
  \bibinfo{author}{\bibfnamefont{N.~M.} \bibnamefont{Shnerb}},
  \bibnamefont{and} \bibinfo{author}{\bibfnamefont{D.~A.}
  \bibnamefont{Kessler}}, \bibinfo{journal}{J. Theor. Biol.}
  \textbf{\bibinfo{volume}{262}}, \bibinfo{pages}{245} (\bibinfo{year}{2009}).

\bibitem[{\citenamefont{Barab{\'a}si and Albert}(1999)}]{barabasi}
\bibinfo{author}{\bibfnamefont{A.-L.} \bibnamefont{Barab{\'a}si}}
  \bibnamefont{and} \bibinfo{author}{\bibfnamefont{R.}~\bibnamefont{Albert}},
  \bibinfo{journal}{Science} \textbf{\bibinfo{volume}{286}},
  \bibinfo{pages}{509} (\bibinfo{year}{1999}).

\bibitem[{\citenamefont{Simon and Bonini}(1958)}]{simon58}
\bibinfo{author}{\bibfnamefont{H.~A.} \bibnamefont{Simon}} \bibnamefont{and}
  \bibinfo{author}{\bibfnamefont{C.~P.} \bibnamefont{Bonini}},
  \bibinfo{journal}{The American Economic Review}
  \textbf{\bibinfo{volume}{48}}, \bibinfo{pages}{607} (\bibinfo{year}{1958}).

\bibitem[{\citenamefont{Clauset et~al.}(2007)\citenamefont{Clauset, Young, and
  Gleditsch}}]{terror1}
\bibinfo{author}{\bibfnamefont{A.}~\bibnamefont{Clauset}},
  \bibinfo{author}{\bibfnamefont{M.}~\bibnamefont{Young}}, \bibnamefont{and}
  \bibinfo{author}{\bibfnamefont{K.~S.} \bibnamefont{Gleditsch}},
  \bibinfo{journal}{Journal of Conflict Resolution}
  \textbf{\bibinfo{volume}{51}}, \bibinfo{pages}{58} (\bibinfo{year}{2007}).

\bibitem[{\citenamefont{Bisby et~al.}(2009)\citenamefont{Bisby, Roskov, Orrell,
  Nicolson, Paglinawan, Bailly, Kirk, Bourgoin, and Baillargeon}}]{COL}
\bibinfo{author}{\bibfnamefont{F.~A.} \bibnamefont{Bisby}},
  \bibinfo{author}{\bibfnamefont{Y.~R.} \bibnamefont{Roskov}},
  \bibinfo{author}{\bibfnamefont{T.~M.} \bibnamefont{Orrell}},
  \bibinfo{author}{\bibfnamefont{D.}~\bibnamefont{Nicolson}},
  \bibinfo{author}{\bibfnamefont{L.~E.} \bibnamefont{Paglinawan}},
  \bibinfo{author}{\bibfnamefont{N.}~\bibnamefont{Bailly}},
  \bibinfo{author}{\bibfnamefont{P.~M.} \bibnamefont{Kirk}},
  \bibinfo{author}{\bibfnamefont{T.}~\bibnamefont{Bourgoin}}, \bibnamefont{and}
  \bibinfo{author}{\bibfnamefont{G.}~\bibnamefont{Baillargeon}},
  \emph{\bibinfo{title}{Species 2000 \& ITIS Catalogue of Life: 2009 Annual
  Checklist. CD-ROM; Species 2000}} (\bibinfo{publisher}{WW Norton \& Company},
  \bibinfo{address}{Reading, U.K.}, \bibinfo{year}{2009}).

\bibitem[{\citenamefont{Clauset et~al.}(2009)\citenamefont{Clauset, Shalizi,
  and Newman}}]{newman2}
\bibinfo{author}{\bibfnamefont{A.}~\bibnamefont{Clauset}},
  \bibinfo{author}{\bibfnamefont{C.~R.} \bibnamefont{Shalizi}},
  \bibnamefont{and} \bibinfo{author}{\bibfnamefont{M.~E.~J.}
  \bibnamefont{Newman}}, \bibinfo{journal}{SIAM Review}
  \textbf{\bibinfo{volume}{51}}, \bibinfo{pages}{661} (\bibinfo{year}{2009}).

\bibitem[{\citenamefont{Broder et~al.}(2000)\citenamefont{Broder, Kumar,
  Maghoul, Raghavan, Rajagopalan, Stata, Tomkins, and Wiener}}]{Broder}
\bibinfo{author}{\bibfnamefont{A.}~\bibnamefont{Broder}},
  \bibinfo{author}{\bibfnamefont{R.}~\bibnamefont{Kumar}},
  \bibinfo{author}{\bibfnamefont{F.}~\bibnamefont{Maghoul}},
  \bibinfo{author}{\bibfnamefont{P.}~\bibnamefont{Raghavan}},
  \bibinfo{author}{\bibfnamefont{S.}~\bibnamefont{Rajagopalan}},
  \bibinfo{author}{\bibfnamefont{R.}~\bibnamefont{Stata}},
  \bibinfo{author}{\bibfnamefont{A.}~\bibnamefont{Tomkins}}, \bibnamefont{and}
  \bibinfo{author}{\bibfnamefont{J.}~\bibnamefont{Wiener}},
  \bibinfo{journal}{Computer Networks} \textbf{\bibinfo{volume}{33}},
  \bibinfo{pages}{309 } (\bibinfo{year}{2000}), ISSN \bibinfo{issn}{1389-1286}.

\bibitem[{\citenamefont{Hubbell et~al.}(2005)\citenamefont{Hubbell, Condit, and
  Foster}}]{hubbella}
\bibinfo{author}{\bibfnamefont{S.~P.} \bibnamefont{Hubbell}},
  \bibinfo{author}{\bibfnamefont{R.}~\bibnamefont{Condit}}, \bibnamefont{and}
  \bibinfo{author}{\bibfnamefont{R.~B.} \bibnamefont{Foster}}
  (\bibinfo{year}{2005}),
  \urlprefix\url{https://ctfs.arnarb.harvard.edu/webatlas/datasets/bci}.

\bibitem[{\citenamefont{Karev et~al.}(2002)\citenamefont{Karev, Wolf, Rzhetsky,
  Berezovskaya, and Koonin}}]{Karev}
\bibinfo{author}{\bibfnamefont{G.}~\bibnamefont{Karev}},
  \bibinfo{author}{\bibfnamefont{Y.}~\bibnamefont{Wolf}},
  \bibinfo{author}{\bibfnamefont{A.}~\bibnamefont{Rzhetsky}},
  \bibinfo{author}{\bibfnamefont{F.}~\bibnamefont{Berezovskaya}},
  \bibnamefont{and} \bibinfo{author}{\bibfnamefont{E.}~\bibnamefont{Koonin}},
  \bibinfo{journal}{BMC Evolutionary Biology} \textbf{\bibinfo{volume}{2}},
  \bibinfo{pages}{18} (\bibinfo{year}{2002}), ISSN \bibinfo{issn}{1471-2148}.

\bibitem[{\citenamefont{Volkov et~al.}(2003)\citenamefont{Volkov, Banavar,
  Hubbell, and Maritan}}]{maritan}
\bibinfo{author}{\bibfnamefont{I.}~\bibnamefont{Volkov}},
  \bibinfo{author}{\bibfnamefont{J.~R.} \bibnamefont{Banavar}},
  \bibinfo{author}{\bibfnamefont{S.~P.} \bibnamefont{Hubbell}},
  \bibnamefont{and} \bibinfo{author}{\bibfnamefont{A.}~\bibnamefont{Maritan}},
  \bibinfo{journal}{Nature} \textbf{\bibinfo{volume}{424}},
  \bibinfo{pages}{1035} (\bibinfo{year}{2003}).

\bibitem[{\citenamefont{Abramowitz and Stegun}(1964)}]{Abramowitz}
\bibinfo{editor}{\bibfnamefont{M.}~\bibnamefont{Abramowitz}} \bibnamefont{and}
  \bibinfo{editor}{\bibfnamefont{I.~A.} \bibnamefont{Stegun}}, eds.,
  \emph{\bibinfo{title}{Handbook of Mathematical Functions with Formulas,
  Graphs, and Mathematical Tables}}, no.~\bibinfo{number}{55} in
  \bibinfo{series}{National Bureau of Standards Applied Mathematics Series}
  (\bibinfo{publisher}{Gov't. Printing Office}, \bibinfo{address}{Washington,
  DC}, \bibinfo{year}{1964}), \bibinfo{edition}{10th} ed.

\bibitem[{\citenamefont{Kessler and Shnerb}(2007)}]{WKB}
\bibinfo{author}{\bibfnamefont{D.~A.} \bibnamefont{Kessler}} \bibnamefont{and}
  \bibinfo{author}{\bibfnamefont{N.~M.} \bibnamefont{Shnerb}},
  \bibinfo{journal}{Journal of Statistical Physics}
  \textbf{\bibinfo{volume}{127}}, \bibinfo{pages}{861} (\bibinfo{year}{2007}).

\bibitem[{\citenamefont{Bohorquez et~al.}(2009)\citenamefont{Bohorquez,
  Gourley, Dixon, Spagat, and Johnson}}]{terror}
\bibinfo{author}{\bibfnamefont{J.~C.} \bibnamefont{Bohorquez}},
  \bibinfo{author}{\bibfnamefont{S.}~\bibnamefont{Gourley}},
  \bibinfo{author}{\bibfnamefont{A.~R.} \bibnamefont{Dixon}},
  \bibinfo{author}{\bibfnamefont{M.}~\bibnamefont{Spagat}}, \bibnamefont{and}
  \bibinfo{author}{\bibfnamefont{N.~F.} \bibnamefont{Johnson}},
  \bibinfo{journal}{Nature} \textbf{\bibinfo{volume}{462}},
  \bibinfo{pages}{911} (\bibinfo{year}{2009}).

\bibitem[{\citenamefont{{Ferrer-i-Cancho} and Servedio}(2005)}]{vito}
\bibinfo{author}{\bibfnamefont{R.}~\bibnamefont{{Ferrer-i-Cancho}}}
  \bibnamefont{and} \bibinfo{author}{\bibfnamefont{V.~D.~P.}
  \bibnamefont{Servedio}}, \bibinfo{journal}{Glottometrics}
  \textbf{\bibinfo{volume}{11}}, \bibinfo{pages}{1} (\bibinfo{year}{2005}).

\bibitem[{\citenamefont{Ricklefs}(2007)}]{ricklefs}
\bibinfo{author}{\bibfnamefont{R.~E.} \bibnamefont{Ricklefs}},
  \bibinfo{journal}{Trends Ecol Evol (Amst)} \textbf{\bibinfo{volume}{22}},
  \bibinfo{pages}{601} (\bibinfo{year}{2007}).

\bibitem{jtb}
Maruvka, Y.~E., Shnerb, N.~M., and Kessler, D.~A.
\newblock {\em J. Theor. Biol.}{ \bf 262}, 245--256 (2009).

\bibitem{barabasi}
Barab{\'a}si, A.-L. and Albert, R.
\newblock {\em Science}{ \bf 286}(5439), 509--512 October  (1999).

\bibitem{newman1}
Newman, M. E.~J.
\newblock {\em Contemporary Phys.}{ \bf 46}(5), 323--351 Sept./Oct.  (2005).

\bibitem{Sta}
Wiechmann, D. and Fuhs, S.
\newblock {\em Corpus Linguistics and Linguistic Theory}{ \bf 2}(1), 107--127
  (2006).

\bibitem{ricklefs2006}
Ricklefs, R.~E.
\newblock {\em Ecology}{ \bf 87}(6), 1424--1431 (2006).

\bibitem{Nee}
Sean, N.
\newblock {\em Functional Ecology}{ \bf 19}(1), 173--176 (2005).

\bibitem{Karev}
Karev, G., Wolf, Y., Rzhetsky, A., Berezovskaya, F., and Koonin, E.
\newblock {\em BMC Evolutionary Biology}{ \bf 2}(1), 18 (2002).

\bibitem{Shakhnovich}
Zeldovich, K.~B., Chen, P., Shakhnovich, B.~E., and Shakhnovich,
E.~I.
\newblock {\em PLoS Comp. Biol.}{ \bf 3}(7), e139: 1224--1238 (2007).

\bibitem{tsallis}
Tsallis, C.
\newblock {\em Journal of Statistical Physics}{ \bf 52}, 479--487 (1988).



\end{thebibliography}
\end{document}